\documentclass[preprint,3p,12pt]{elsarticle}

\usepackage{amsmath,amssymb}
\usepackage{mathtools}
\usepackage{bm}
\usepackage{graphicx}
\usepackage{MnSymbol}

\usepackage{fancyhdr}

\biboptions{numbers,sort&compress}

\usepackage{hyperref}

\begin{document}

\begin{frontmatter}

\title{Identification of perturbative ambiguity canceled against bion}

\author{Okuto Morikawa}
\ead{o-morikawa@phys.kyushu-u.ac.jp}
\author{Hiromasa Takaura}
\ead{takaura@phys.kyushu-u.ac.jp}
\address{Department of Physics, Kyushu University,
744 Motooka, Nishi-ku, Fukuoka, 819-0395, Japan}

\begin{abstract}
It was conjectured that bions,
semi-classical objects found in a compactified spacetime,
are responsible for the cancellation of the so-called renormalon ambiguities.
Contrary to the conjecture, we argue that the ambiguity due to the bion
corresponds to the proliferation of Feynman diagrams.
We point out that the amplitudes of almost all Feynman diagrams are enhanced
due to modifications of the infrared structure of perturbation theory
upon an $S^1$ compactification and twisted boundary conditions.
Our findings clarify the role of the semi-classical object in resurgence structure,
which has been a controversial issue in recent years.
\end{abstract}

\end{frontmatter}

\thispagestyle{fancy}
\rhead{KYUSHU-HET-210}
\renewcommand{\headrulewidth}{0.0pt}

\section{Introduction and outline}

In quantum field theory, perturbation theory is extensively used
as a basic and general method to analyze phenomena in particle physics.
However, it is known that perturbation theory often possesses
intrinsic errors in its predictions.
This is caused by divergent behaviors of
perturbation series, $\sum_{k} a_k [g^2/(16 \pi^2)]^k$,
due to factorial growth of perturbative coefficients $a_k$ at large orders.
This poses a fundamental problem how one can achieve ultimate predictions.
Recently, the so-called resurgence structure is believed where 
the perturbative ambiguities are eventually canceled
against ambiguities of nonperturbative calculations, and there are vigorous discussions
on its concrete structure~\cite{Dunne:2002rq,Dunne:2015eaa}.

It has been recognized that there are two origins of
the factorial growth of the perturbative coefficients.
One is caused by the proliferation of Feynman diagrams (PFD),
and the other relates to renormalization properties
and is known as renormalon~\cite{tHooft:1977xjm,Beneke:1998ui}.
They induce factorially growing perturbative coefficients\footnote{%
For $d$-dimensional spacetime, we adopt a convention
$\sum_{k} a_k [g^2/(4 \pi)^{d/2}]^k$.}
and result in perturbative ambiguities, $\delta$, typically as follows:
\begin{align}
 \text{PFD}&: &
 a_k &\sim k! , &
 \delta &\sim e^{-16 \pi^2/g^2} ,
 \notag\\
 \text{Renormalon}&: &
 a_k &\sim \beta_0^k k! , &
 \delta &\sim e^{-16 \pi^2/(\beta_0 g^2)} ,
\end{align}
where $\beta_0$ is the one-loop coefficient of the beta function
[e.g. $\beta_0=11 N/3$ for $SU(N)$ quenched QCD].
One can see that the perturbative ambiguities reduce to nonperturbative factors.
It is convenient to quantitatively characterize the perturbative ambiguities
by singularities of the Borel transform $B(t)=\sum_{k=0}^{\infty} (a_k/k!) t^k$,
the generating function of the perturbative coefficients.
The PFD induces the singularities at $t=1$, $2$, \dots,
while the renormalon at~$t=1/\beta_0$, $2/\beta_0$, \dots.
The PFD-type ambiguity is known to get canceled against
the ambiguity of the semi-classical calculation
for the instanton--anti-instanton amplitude
$\sim e^{-2 S_{I}}$~\cite{Bogomolny:1980ur,ZinnJustin:1981dx},
where $S_{I}= 8\pi^2/g^2$ is the one-instanton action.
On the other hand, the cancellation of a renormalon ambiguity was shown
only in a two dimensional non-linear sigma model~\cite{David:1982qv}.

In 2012, the conjecture
was proposed~\cite{Argyres:2012vv,Argyres:2012ka,Dunne:2012ae,Dunne:2012zk}
that renormalon ambiguities are canceled
by the new semi-classical object called a bion~\cite{Unsal:2007jx},
a pair of a fractional instanton and anti-fractional instanton.
This conjecture first requires an $S^1$ compactification of a spacetime as
$\mathbb{R}^d\to\mathbb{R}^{d-1}\times S^1$
with a small $S^1$-radius $R$ such that $R\Lambda\ll 1$,
where $\Lambda$ is a dynamical scale of an asymptotically free theory.
A bion solution appears in this setup when twisted boundary 
conditions are imposed along the $S^1$-direction 
(or equivalently when a non-trivial holonomy exists under the periodic boundary condition).
The bion action is given by $S_B=S_I/N$,
where $N$ is a parameter specifying the degree of freedom of dynamical variables.
Accordingly, the ambiguities in bion calculus are typically given by~$\sim e^{-2S_I/N}$.
The corresponding perturbative ambiguities are specified
by the Borel singularities at $t=1/N$, $2/N$, \dots.
Due to similar $N$ dependence to renormalon ambiguities,
it was conjectured that the bion is responsible
for the cancellation of renormalon ambiguities.
Since then, active discussions on the conjecture have been initiated.

However, recent studies have reported inconsistency of the conjecture.
Two observations were made in the systems where bion ambiguities exist.
(i) In the $SU(2)$~and~$SU(3)$ gauge theories
with adjoint fermions on~$\mathbb{R}^3\times S^1$,
renormalon ambiguities are absent~\cite{Anber:2014sda}.
(ii)~In the $\mathbb{C}P^{N-1}$ models on $\mathbb{R}\times S^1$,
renormalon ambiguities are specified by the Borel singularity 
at $t=3/(2N)$~\cite{Ishikawa:2019tnw,Ishikawa:2019oga},
which conflicts with that of the bion ambiguity~\cite{%
Fujimori:2016ljw,Fujimori:2017oab,Fujimori:2017osz,Fujimori:2018kqp}.\footnote{%
The renormalon analyses~\cite{Ishikawa:2019tnw,Ishikawa:2019oga}
have been performed for $N R\Lambda\gg 1$,
in which the bion analyses are not always valid.
An important point is that it was shown~\cite{Ishikawa:2019tnw,Ishikawa:2019oga}
that the $S^1$ compactification affects renormalon structure
even in this so-called large-$N$ volume independence domain.
This is against an expectation of the conjecture.}
The both cases indicate that the bion ambiguities
do not correspond to the renormalon ambiguities.

These observations give rise to a new question:
what cancels bion ambiguities?
If renormalon does not correspond to bions,
there should be a different type of perturbative ambiguities
which cancel bion ambiguities.
Such a perturbative ambiguity has not been identified so far.
This implies a lack of our understanding on the resurgence structure.

In this Letter, we argue that bion ambiguities are canceled
against PFD-type ambiguities.
Although this possibility was mentioned in Ref.~\cite{Anber:2014sda},
so far, it has not been understood whether a seemingly different
singularity (by the factor $1/N$) can get compatible with the bion ambiguity.
We clarify that it occurs by a non-trivial enhancement
of the amplitudes of almost all Feynman diagrams
as a consequence of the $S^1$ compactification and twisted boundary conditions.
As a result, an $N$ times closer Borel singularity to the origin
arises consistent with the bion.
Our findings are helpful in giving a unified understanding to a series of
controversial discussions.

We explain our setup.
We consider $\mathbb{R}^{d-1}\times S^1$ spacetime with the $S^1$-radius~$R$,
and the $\mathbb{Z}_N$-twisted boundary conditions 
for an $N$-component field~$\phi^A$
\begin{equation}
 \phi^A (\bm{x}, x_d+2\pi R) = e^{i m_A 2\pi R} \phi^A(\bm{x}, x_d) .
 \label{eq:twist_bc}
\end{equation}
Here $\bm{x}$ denotes the coordinates~$(x_1,\dots,x_{d-1})$ of~$\mathbb{R}^{d-1}$,
$x_d$ does that of~$S^1$, and the twist angle $m_A$ is given by
\begin{equation}
 m_A =
 \begin{cases}
  A/(N R) & \text{for $A=1$, $2$, \dots, $N-1$,}\\
  0       & \text{for $A=N$.}
 \end{cases}
 \label{eq:twist_mass}
\end{equation}
This setup allows us to find a bion solution 
with the bion action $S_B=(4 \pi)^{d/2} m_A R/g^2$ with~$A<N$,
and the bion ambiguities $\sim e^{- S_B}$ appear.
The corresponding perturbative ambiguities
are specified by the Borel singularities at $t=1/(m_A R)=A/N=1/N$, $2/N$, \dots.

We explain the keys to understanding the enhancement of 
the PFD-type ambiguity under the above setup.
First, the $S^1$ compactification reduces
the dimension of a momentum integral from~$d$ to~$d-1$.
Then, perturbation theory tends to suffer from severe infrared (IR) divergences.
This is parallel to
finite-temperature and lower-dimensional super-renormalizable field theories
with massless particles~\cite{Gross:1980br,Kapusta:2006pm}.
Secondly, the twist angles play the role of mass terms
for the zero Kaluza--Klein (KK) momentum, and hence work as IR regulators.
As a result, an amplitude of a Feynman diagram is given 
by an inverse power of the twist angle as~$[g^2/(m_A R)]^k$ at the $k$th order,
as a signal of the IR divergence. (It diverges when $m_A \to 0$.)  
For small $A$, this behavior gives $\sim (N g^2)^k$ due to Eq.~\eqref{eq:twist_mass}.
Thus, if we have the PFD exhibiting the IR divergences in the above sense,
the $k$th order term of the perturbation series behaves as $\sim k!(N g^2)^k$
rather than $\sim k!(g^2)^k$.
This causes the Borel singularities at $t=1/N$, $2/N$, \dots,
corresponding to the bion ambiguities.

In the following, we study the $\mathbb{C}P^N$ model on~$\mathbb{R}\times S^1$
as a concrete model, where an explicit calculation of the bion ambiguities
has been performed~\cite{Fujimori:2018kqp}.

\section{Enhancement of perturbative ambiguity: the $\mathbb{C}P^{N-1}$ model}

The $\mathbb{C}P^{N-1}$ model in the two-dimensional Euclidean spacetime 
is given by
\begin{equation}
 S = \frac{1}{g^2} \int d^2 x\,
 \left[\partial_\mu\Bar{z}^A\partial_\mu z^A - j_\mu j_\mu
 + f(\Bar{z}^A z^A - 1) \right],
\end{equation}
in terms of homogeneous coordinates $z^A$~and~$\Bar{z}^A$,
where $f$ is introduced as the Lagrange multiplier field
to impose the constraint~$\Bar{z}^A z^A=1$, and
\begin{equation}
 j_\mu \equiv \frac{1}{2i} \Bar{z}^A\overleftrightarrow\partial_\mu z^A ,
  \qquad
 \overleftrightarrow\partial_\mu \equiv \partial_\mu - \overleftarrow\partial_\mu .
\end{equation}
We implement the $\mathbb{Z}_N$-twisted boundary conditions~\eqref{eq:twist_bc}
for $z^A$ along the $x_2$-direction.
Then the propagator of~$z^A$ is obtained as
\begin{align}
 \left\langle z^A(x) \Bar{z}^B(y) \right\rangle
 &=g^2 \delta^{AB} \sumint_p
 \frac{e^{i(p_\mu+m_A\delta_{\mu2})(x-y)_\mu}}{p_1^2+(p_2+m_A)^2+f_0}
 \notag\\
 &\equiv g^2 \delta^{AB}\square^{-1}_A(x-y) , \label{eq:propagator}
\end{align}
with the vacuum expectation value $f_0$, which is determined by the tadpole condition
that the linear term of the fluctuation $\delta f$ vanishes.
Here and hereafter, we use the abbreviation,
\begin{equation}
 \sumint_p \equiv \int\frac{d p_1}{2\pi} \frac{1}{2\pi R}\sum_{p_2\in\mathbb{Z}/R} .
\end{equation}
Note that, in the denominator of~Eq.~\eqref{eq:propagator},
$m_A^2+f_0$ serves as a mass parameter for $p_2=0$.

We consider the partition function in perturbation theory,
\begin{align}
 \mathcal{Z}
 &= \int\mathcal{D}f \int\mathcal{D}z^A\mathcal{D}\Bar{z}^A\, e^{-S}
 \notag\\
 &= \int\mathcal{D}f\, e^{(1/g^2)\int d^2x\, f} \mathcal{Z}'[f]
\end{align}
where
\begin{align}
 &\mathcal{Z}'[f=f_0+\delta f]
 \notag\\
 &\equiv
 \exp\left(g^2 \frac{\delta}{\delta\Bar{z}^A}\cdot\square^{-1}_A\cdot
 \frac{\delta}{\delta z^A}\right)
 \notag\\&\qquad\times
 \left.
 \exp\left[\frac{1}{g^2}
 \int d^2 x\, \left(j_\mu j_\mu- \delta f \bar{z}^A z^A\right)\right]
 \right|_{z=0} .
\end{align}
In the following, we consider only the $A=1$ flavor.
As we shall see, this contribution is significant for enhancement of the amplitude.
(Note that a flavor symmetry is broken by the twisted boundary conditions.)
We study the $\mathcal{O}(\delta f^0)$ term in $\mathcal{Z}'[f]$ for simplicity.

We first give a review on the PFD in the non-compactified case. 
We estimate the $k$th order perturbative contribution
by equally assigning one to every possible diagram.
For this purpose, it is suitable to consider
the replacements $\square_A^{-1}\to 1$ and~$j_\mu\to\Bar{z}z$.
This approximately corresponds to employing the zero-dimensional version of the model.
Then $\mathcal{Z}'[f]\equiv\sum_{k=0}^{\infty}T_k(g^2)^k$ is estimated as
\begin{align}
 T_k &\sim \frac{1}{(2k)!}
 \left(\frac{\delta}{\delta\Bar{z}}\frac{\delta}{\delta z}\right)^{2k}
 \frac{1}{k!}\left[\left(\Bar{z} z\right)^2\right]^k
 \notag\\
 &= \frac{(2 k)!}{k!} \sim 4^k \Gamma(k+1/2) .
 \label{eq:Tk}
\end{align}
This factorial growth is interpreted as the source of
the Borel singularities at $t=1$, $2$, \dots in the non-compactified spacetime,
corresponding to the ambiguities
in the instanton--anti-instanton calculus~\cite{Bogomolny:1980ur,ZinnJustin:1981dx}.

Now we explain how the enhancement in terms of~$N$ occurs
under the $S^1$ compactification and twisted boundary conditions.
For this purpose, the above naive counting is not sufficient
and we need to have a closer look at the structure of loop integrals.
The $k$th order perturbative contribution to $\mathcal{Z}'[f]$ is given by
\begin{align}
 &\mathcal{Z}'[f]|_{\text{$k$th order}}
 \notag\\
 &= \frac{1}{(2k)!}
 \left(g^2 \frac{\delta}{\delta \bar{z}^A}\cdot\square^{-1}_{A}\cdot
 \frac{\delta}{\delta z^A} \right)^{2k}
 \notag\\&\qquad\times
 \frac{1}{k!} \left[\frac{1}{g^2} \left(-\frac{1}{4}\right)
 \int d^2 x \left(\Bar{z}^A \overleftrightarrow\partial_\mu z^A\right)^2 \right]^k.
\end{align}
The number of vertices, $V$, is identical to the perturbation order, $V=k$.
We first consider connected Feynman diagrams,
for which the following relations follow:
\begin{align}
 P &= 2V ,& L &= V + 1, \label{eq:diagramrelations}
\end{align}
where $P$~and~$L$ denote the numbers of propagators and loops, respectively.
Noting these relations, an amplitude of a Feynman diagram is written as
\begin{equation}
 V_2 (g^2)^k \sumint_{p_1, \dots, p_{k+1}} 
 \frac{F^{(2k)}(p_{i, \mu}+m_A \delta_{\mu 2})}
  {\prod_{i=1}^{2k} [q_{i,1}^2+(q_{i,2}+m_A)^2+f_0]} ,
\end{equation}
where $V_2$ is the volume factor of the two-dimensional spacetime,
$p_i$ denotes a loop momentum,
and $q_i$~is the momentum of a propagator,
which is given by a linear combination of ($p_1$, \dots, $p_{k+1}$).
In the numerator, we have a $(2k)$th-order homogeneous polynomial $F^{(2k)}$,
originating from the derivative in the interaction term.

The enhancement of the amplitude is caused by the zero KK-modes
where~$p_{i,2}=0$ for~${}^\forall i$, and hence, $q_{i,2}=0$ for ${}^\forall i$.
For this part, we have
\begin{equation}
 \frac{V_2 (g^2)^k}{(2\pi R)^{k+1}}
 \int\left(\prod_{i=1}^{k+1}\frac{d p_{i,1}}{2\pi}\right)
 \frac{F^{(2k)}(p_{i,1},m_A)}{\prod_{i=1}^{2k} [q_{i,1}^2+m_A^2+f_0]} .
 \label{eq:loopintegral}
\end{equation}
Note that each loop integral becomes a one-dimensional integral due to the $S^1$ compactification.
Accordingly, this expression possesses an IR divergence
in the massless limit $m_A^2+f\to0$ with the degree of divergence $k-1$;
the mass dimension of the integration measures is $(k+1)$, 
whereas that of the integrand is $-2k$.
Then, if $f_0$ is negligible compared with $m_{A=1}=1/(NR)$,
we obtain
\begin{equation}
 \sim \frac{V_2}{R^2} \frac{1}{(m_A R)^{k-1}} \left(\frac{g^2}{4\pi}\right)^k 
 =\frac{V_2}{R^2} \frac{1}{N} \left(\frac{N g^2}{4 \pi}\right)^k
 \label{eq:connectedcontribution}
\end{equation}
because $m_A$ works as an IR regulator. 
(See below for discussion on the size of $f_0/m_A^2$.)
We note that this argument indicates that a \textit{general} connected diagram 
has the contribution naturally specified by $N^k g^{2k}$ rather than~$g^{2k}$.

We see that disconnected diagrams show weaker enhancement.
As an example, let us consider a $k$th order disconnected diagram 
given as a product of two connected diagrams.
Since each connected diagram satisfies Eq.~\eqref{eq:diagramrelations},
the amplitude is given by
\begin{equation}
 \sim \left(\frac{V_2}{R^2}\right)^2  \frac{1}{N^2} \left(\frac{N g^2}{4\pi}\right)^{k} .
\end{equation}
from Eq.~\eqref{eq:connectedcontribution}.
(Note that the total number of vertices is fixed as $k$.)
This is suppressed by the inverse power of $N$
compared with the contribution of the connected diagram~\eqref{eq:connectedcontribution}.

\begin{figure}[t]
 \begin{center}
  \includegraphics[width=0.7\columnwidth]{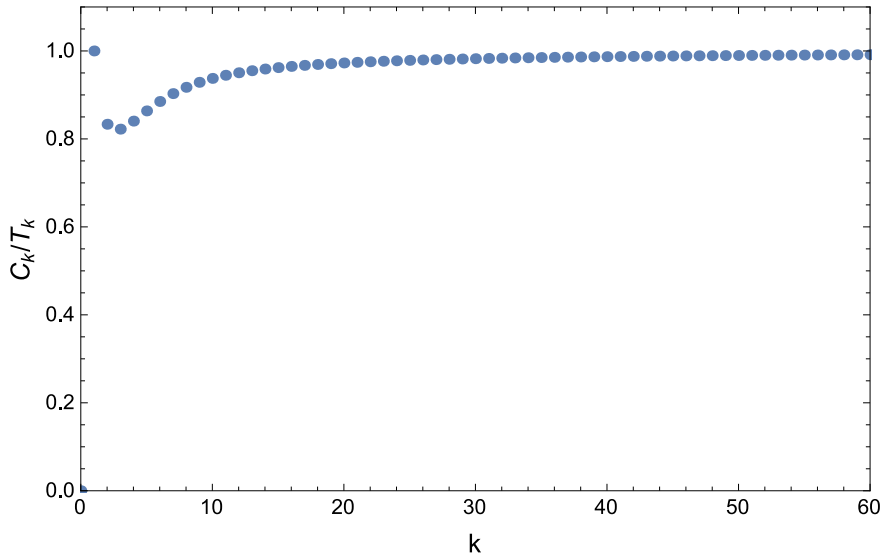}
  \caption{Estimate of $C_k/T_k$.}
  \label{fig:CkoverTk}
 \end{center}
\end{figure}

We can show that connected diagrams,
which have the strongest enhancement, increase factorially.
To show this, it is sufficient to assign one to every connected diagram.
One can use
\begin{equation}
 C=\sum_{k=0}^{\infty} C_k (g^2)^k = \ln\left[\sum_{k=0}^{\infty}T_k(g^2)^k\right] ,
\end{equation}
since the partition function $\mathcal{Z}'$ is given by
the exponential of the total sum of connected diagrams $C$ (i.e. $\mathcal{Z}'=e^C$).
From the estimate of $T_k$ in Eq.~\eqref{eq:Tk} and this formula,
we estimate $C_k$ in Fig.~\ref{fig:CkoverTk}.
One sees that $C_k$ grows factorially at the almost same rate as $T_k$.
Thus, it is plausible that we have the perturbation series
as $\sum_{k} k! (N g^2)^k$
and the Borel singularity at $t=1/N$, corresponding to the bion ambiguity.

We emphasize that both the bion ambiguity and the position of the Borel singularity
are uniformly controlled by the twist angles. 
This indicates validity of the resurgence structure between the bion 
and enhanced perturbative contributions.

We note that the above enhancement is specific to the case of $m_A^2 \gg f_0$.
In contrast, if $m_A^2\ll f_0$,
since $f_0$ works as an IR regulator, we have
\begin{align}
 \sim\frac{V_2}{R^2} \sum_{\alpha \geq 0} c_{\alpha}
 \frac{(m_A R)^\alpha}{(f_0 R^2)^{(k+\alpha-1)/2}}
 \left(\frac{g^2}{4\pi}\right)^k .
 \label{eq:linde}
\end{align}
The enhancement does not occur in this case.

The size of $m_A^2/f_0$,
which is critical for the existence of the $N$ times closer Borel singularity,
is determined depending on the magnitude of  $N R\Lambda$.
To determine the vacuum expectation value $f_0$,
we consider the tadpole condition at the one-loop level.
For $N R\Lambda\ll 1$, one finds a perturbative solution 
$\sqrt{f_0}R\sim g^2((NR)^{-1})/(4 \pi)$
in terms of the renormalized coupling~$g^2(\mu)$~\cite{Yamazaki:2019arj}.
Hence $m_A^2 \gg f_0$ is satisfied within the perturbative expansion.
Thus, the Borel singularity at $t=1/N$ is likely to arise. 
It is worth noting that $N R\Lambda\ll 1$ also ensures validity of bion analyses.
This shows overall consistency.

On the other hand, for $N R\Lambda\gg 1$,
$f_0$ is given by $f_0=\Lambda^2$ as in the non-compactified case
due to the so-called large~$N$ volume independence~\cite{Sulejmanpasic:2016llc}.
Then since $m_A^2/f_0=1/(N R\Lambda)^2\ll 1$,
we do not have the above enhancement or a Borel singularity at $t=1/N$
from the PFD.
In fact, since we have inverse powers of $\Lambda$ in this case \cite{Ishikawa:2020eht},
genuine perturbative analysis cannot be considered reasonably.
For $N R\Lambda\gg 1$, the bion analyses are not always valid~\cite{%
Argyres:2012vv,Argyres:2012ka,Dunne:2012ae,Dunne:2012zk,%
Fujimori:2016ljw,Fujimori:2017oab,Fujimori:2017osz,Fujimori:2018kqp}.

\begin{figure}[t]
 \begin{center}
  \includegraphics[width=0.8\columnwidth]{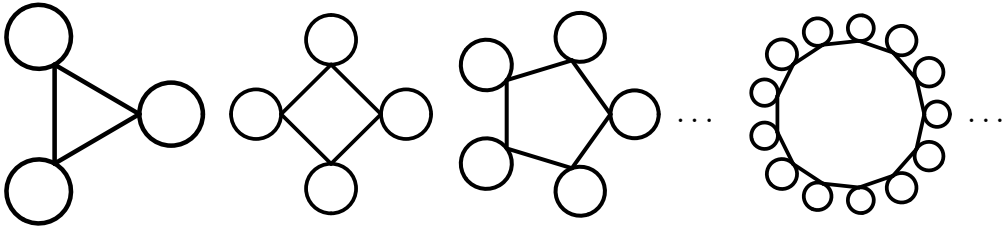}
  \caption{A series of Feynman diagrams
  whose amplitude at the $k$th order is of order~$N^{2k-1} g^{2k}$.}
  \label{fig:bubble}
 \end{center}
\end{figure}

We give supplementary explanations in the case $N R\Lambda\ll 1$,
where we have the enhancement.
First, we mention effects of other flavor contributions than $A=1$. 
From the zero KK modes, we have 
$k!/(m_A R)^{k-1}[g^2/(4 \pi)]^k\sim k!(N/A)^{k-1}[g^2/(4 \pi)]^k$
when we consider the $A$ flavor alone.
Here, $A/N$ corresponds to the position of the Borel singularity.
Then, the closest singularity is given by $A=1$,
and so is the asymptotic behavior of the perturbative coefficients. 
This is the reason why we focused on this contribution.
Secondly, we clarify difference
in power counting of $N$ from the non-compactified case.
So far, we have not considered sums over flavor indices and limited the flavor to $A=1$.
Nevertheless, $N$ dependence appears as a consequence of loop momentum integrations,
where, in particular, the IR structure of loop integrands is essential.
This is in contrast to the non-compactified case, where $N$ dependence arises
exclusively from the sums over flavor indices,
and momentum integrations are irrelevant.
When one takes into account sums over indices as well in the compactified case,
one needs to pay attention to both 
the degree of IR divergence and structure of flavor indices.
For instance, the $N$ dependence of the diagram depicted in Fig.~\ref{fig:bubble}
is given by $N^{2k-1} g^{2k}$ at the $k$th order,
provided that flavor indices for individual loops are independent.
Here, we note that the loops outside a central loop do not possess IR divergences.
Thirdly, we note that, in fact, the $A=N-1$ sector also has the strongest enhancement.
This is because $p_2+m_A$ can be $-1/(N R)$ for $p_2=-1/R$.
Thus, this part has the same order contribution as the $A=1$ sector.

Although we have considered the $\mathcal{O}(\delta f^0)$ term so far,
we can repeat a similar analysis, for instance, for
the quadratic term in $\delta f$ of the effective action for this field,
which has a clearer physical meaning.

Finally, we make some remarks.
Adopting the homogeneous coordinates as in the present study, 
we actually have a difficulty 
in defining fixed order perturbation theory for the effective action.
In integrating out the $A=N$ flavor, 
the IR regulator is given by $f_0$ since~$m_A=0$.
Namely, $z^N$ is subject to the periodic boundary condition
and  we partially have the same situation as finite-temperature field theory.
For $N R\Lambda\ll 1$, $f_0$ is given by $f_0\sim g^4/(4 \pi R)^2$
and plays the role of a screening mass~\cite{Gross:1980br,Kapusta:2006pm}.
It should be kept in the denominator of Eq.~\eqref{eq:loopintegral},
otherwise calculations break down due to IR divergences.
In this treatment, we have the term with $\alpha=0$ in~Eq.~\eqref{eq:linde},
and the $k$th order perturbative contribution is partially given
by~$g^{2k}/ (f_0 R^2)^{(k-1)/2} = O(g^2)$.
It is irrelevant to the perturbation order $k$.
Therefore, higher-loop diagrams can contribute at the same order.
This kind of problem is known as the Linde problem~\cite{Linde:1978px,Linde:1980ts}.
Then, it is practically impossible to systematically obtain a series expansion
in $g^2$ for the effective action even at relatively low orders.
If we instaed adopt inhomogeneous coordinates $\varphi^a=z^a/z^N$
($a=1$, $2$, \dots, $N-1$), perturbative calculations are free
from the Linde problem because all twist angles are taken non-zero.
With the use of the inhomogeneous coordinates
we can also argue that perturbative coefficients behave as $\sim k! (N g^2)^k$
for the partition function  in a similar manner.
We have used the homogeneous coordinates here for simpler illustration.

\section{Conclusions}

For nearly a decade, there have been vigorous discussions on the conjecture
that the bion is responsible for the cancellation of renormalon ambiguities,
and a unified interpretation has not been established.
In this Letter, we argued convincingly
that the bion cancels the perturbative ambiguity caused by the 
proliferation of Feynman diagrams.
In particular, we demonstrated how a Borel singularity appears at $t=1/N$
upon the $S^1$ compactification and twisted boundary conditions,
which is located at $t=1$ in the non-compactified spacetime.
Here, the IR structure of loop integrals is crucial, and 
the $N$ times closer singularity is regarded as a signal of IR divergences
in perturbation theory.
This observation settles a recent controversy about the role of the bion
and deepens our understanding on the resurgence structure.

\section*{Acknowledgments}
The authors are grateful to H.~Suzuki for encouragement and valuable comments.
They also thank M.~Ishii, E.~Itou, and K.~Yonekura for discussions.
This work was supported by JSPS Grants-in-Aid for Scientific Research numbers
JP18J20935 (O.M.) and JP19K14711 (H.T.).

\bibliographystyle{elsarticle-num-names}
\bibliography{N-enhancement_plb}

\end{document}